\documentclass{egpubl}
\usepackage{egsgp18}

 \SpecialIssuePaper         
 \electronicVersion 

\ifpdf \usepackage[pdftex]{graphicx} \pdfcompresslevel=9
\else \usepackage[dvips]{graphicx} \fi

\PrintedOrElectronic

\usepackage{t1enc,dfadobe}

\usepackage{egweblnk}
\usepackage{cite}
\usepackage{amsgen,amsmath,amstext,amsbsy,amssymb,amsopn,amssymb}

\newtheorem{corollary}{Corollary}
\newtheorem{proposition}{Proposition}

\usepackage{xspace}
\usepackage{dsfont}
\makeatletter
\DeclareRobustCommand\onedot{\futurelet\@let@token\@onedot}
\def\@onedot{\ifx\@let@token.\else.\null\fi\xspace}

\def\etal{\emph{et al}\onedot}
\makeatother

\usepackage[dvipsnames]{xcolor}
\newcommand{\rev}[1]{#1}

\usepackage{booktabs}
\usepackage{todonotes}
\usepackage{graphicx}
\usepackage{makecell}
\usepackage{multirow}

\title[Learning Fuzzy Set Representations of Partial Shapes on Dual Embedding Spaces]
{Learning Fuzzy Set Representations of Partial Shapes\\on Dual Embedding Spaces}

\author[M. Sung \& A. Dubrovina \& V. Kim \& L. Guibas]
{\parbox{\textwidth}{\centering
        Minhyuk Sung$^{1}$
        , Anastasia Dubrovina$^{1}$
        , Vladimir G. Kim$^{2}$
        , and Leonidas Guibas$^{1}$
        }
        \\
{\parbox{\textwidth}{\centering
        $^1$Stanford University\\
        $^2$Adobe Research
        }
}
}

\begin{document}

\teaser{
  \centering
  \vspace{-5mm}
  \includegraphics[width=\linewidth]{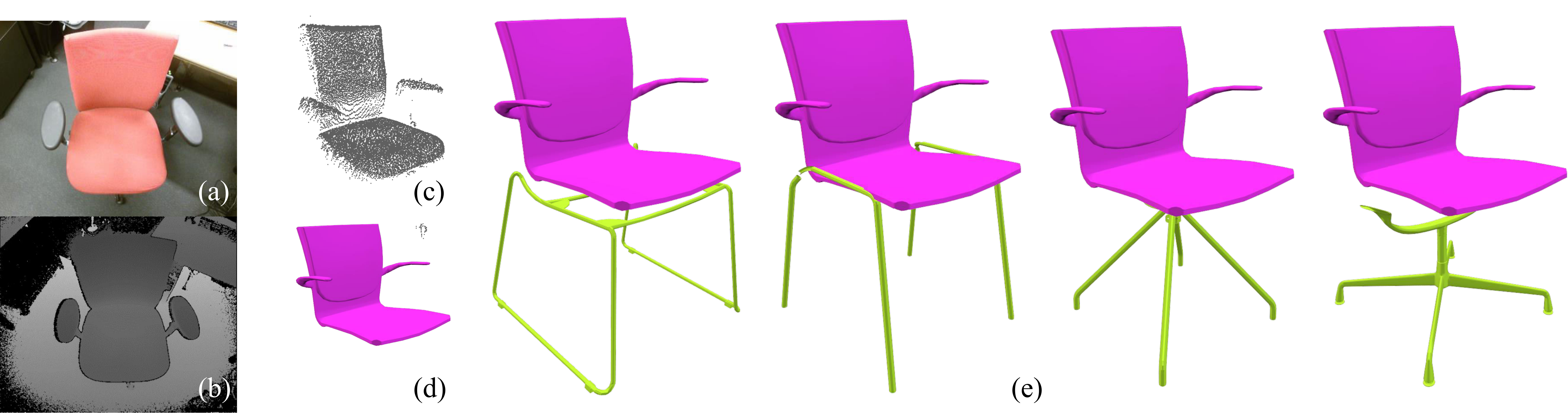}
  \caption{An example of completing a partial scan data with various complement suggestions using our method. Partial objects are represented as fuzzy sets in our embedding, and their complements can be retrieved with a fuzzy set operation in the embedding space. (a) and (b) are a Kinect color and depth image from \cite{Sung:2015}, (c) is a point cloud after removing background, (d) is a partial shape retrieved from our dataset using ICP with a manual initial pose, and (e) are complete objects with four out of the best ten complement retrievals (green parts). The position of complements are automatically computed using a placement network in \cite{Sung:2017}, which is retrained with the partial object data.}
\label{fig:teaser}
}

\maketitle


\begin{abstract}

Modeling relations between components of 3D objects is essential for many geometry editing tasks. Existing techniques commonly rely on labeled components, which requires substantial annotation effort and limits components to a dictionary of pre-defined semantic parts. We propose a novel framework based on neural networks that analyzes an uncurated collection of 3D models from the same category and learns two important types of semantic relations among full and partial shapes: complementarity and interchangeability.  The former helps to identify which two partial shapes make a complete plausible object, and the latter indicates that interchanging two partial shapes from different objects preserves the object plausibility. Our key idea is to jointly encode both relations by embedding partial shapes as fuzzy sets in dual embedding spaces. We model these two relations as fuzzy set operations performed across the dual embedding spaces, and within each space, respectively. We demonstrate the utility of our method for various retrieval tasks that are commonly needed in geometric modeling interfaces.



\end{abstract}


\section{Introduction}
\label{sec:introduction}

Component-based 3D object modeling is common in design of man-made objects. Creating automatic or semi-automatic tools for such component-based modeling has been a long-standing goal in 3D object processing. Towards that goal, previous work leveraged 3D CAD model datasets with known components and component structures for creating new objects \cite{Funkhouser:2004}, completing partial shapes \cite{Shen:2012}, and for analyzing shape structures \cite{Kim:2013a,Li:2017}. Recently created large-scale 3D CAD datasets \cite{Warehouse,Chang:2015} offer significant diversity of component geometry and structure, thus increasing the potential of component-based geometry processing for practical usages.


While a large database increases the diversity in modeling and editing shapes, it also requires the burden of annotating shape components in a consistent manner. Most previous part labeling methods are not scalable \cite{Kaick:2013}, or require some human supervision \cite{Yi:2016}.
To avoid the labor of an annotation, while still taking advantage of the large scale of the databases, Sung \etal \cite{Sung:2017} proposed an annotation-free component assembly method, which trains a neural network to retrieve plausible complementary \emph{components} given a query partial object. The relations among partial objects and complementary components are learned from the data. However, the incremental approach of Sung \etal does not account for the plausibility of the full constructed shape, or can detect \emph{groups of segments} which can plausibly complete a given partial shape. This functionality is important when the plausibility of the full object is important, for example when creating a new object by mixing components from different models in the database \cite{Xu:2012}, or when completing partially created objects \cite{Sung:2015}.

To address these limitations, we propose an annotation-free deep learning framework which learns partial shape representations from database component assemblies, and jointly encodes two semantic relations between partial shapes: \emph{complementarity} and \emph{interchangeability}. Complementarity means that the two partial shapes can be combined into a complete, semantically meaningful object. Interchangeability indicates that replacing a part of a model with another partial shape still produces a plausible new object. This relations can capture semantic similarities among partial shapes in terms of their usage in the context of full objects, even when these partial shapes are geometrically dissimilar. Both complementarity and interchangeability are closely related to each other, since interchangeability means that two partial shapes share the same \rev{set of complements}.

\begin{figure}[t!]
\centering
\includegraphics[width=0.7\linewidth]{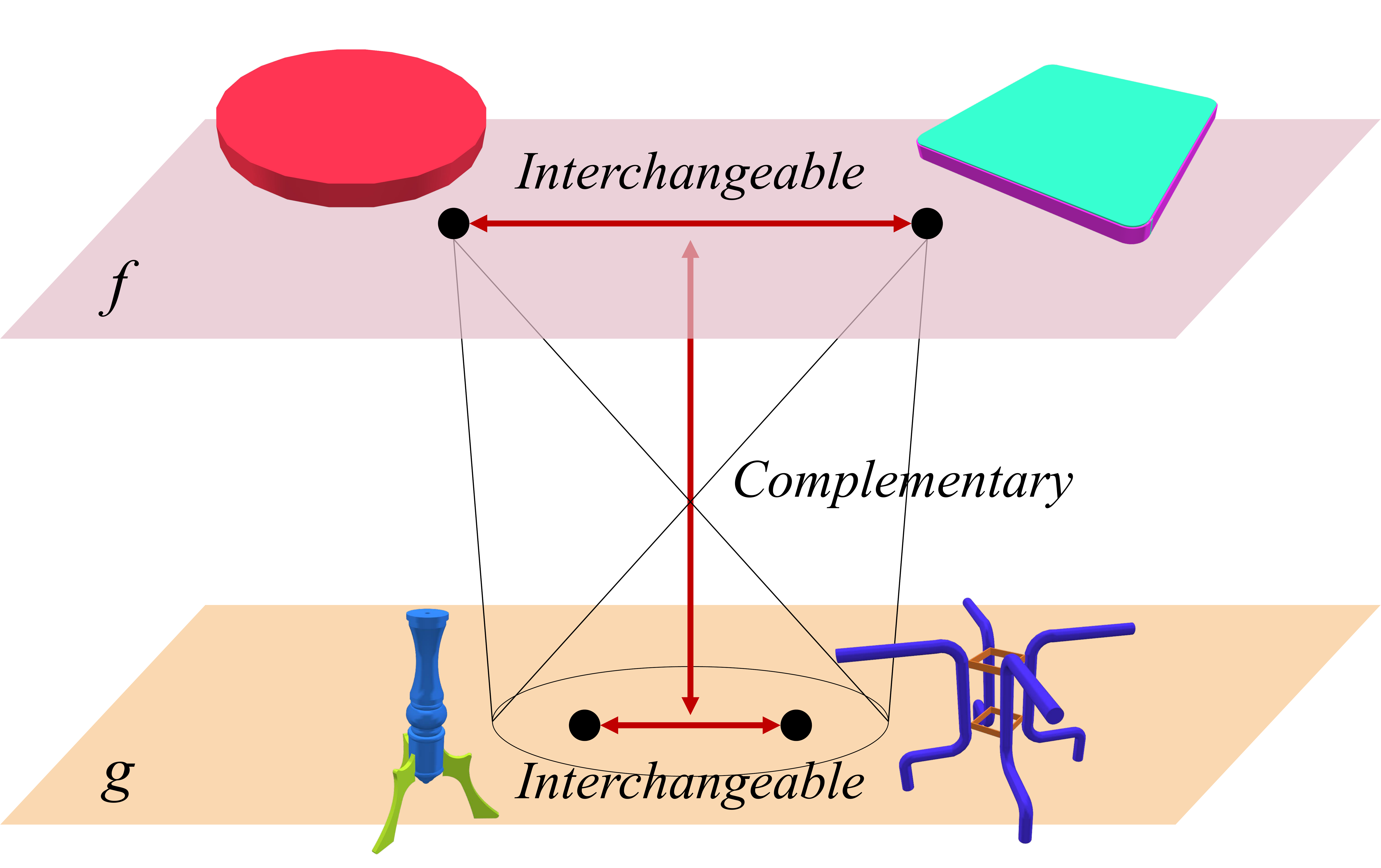} 
\caption{We propose a dual space embedding that encode both complementarity and interchangeability. The complementarity is represented by inter-space relations, and the interchangeability is represented by intra-space relations.}
\label{fig:dual_spaces}
\vspace{-0.5cm}
\end{figure}

Encoding these relations in embedding spaces is not trivial. Complementarity is an \emph{irreflexive} relation, therefore a na\"ive embedding scheme which minimizes distances among related data is not applicable. In addition, we do not have any supervision for learning interchangeability relations, and need to infer this from the complementarity relations between partial shapes. To tackle these challenges, we suggest a novel embedding approach into \emph{dual} embedding spaces \rev{(Figure~\ref{fig:dual_spaces})}. We consider the symmetric (undirected) complementary relation as both-way directed relations, and create two embedding spaces $f$ and $g$ for one-to-$N$ mapping, such that all variations of partial shapes are present in both spaces. Given two partial shapes, complementarity between them is reflected by their embeddings into the two spaces.
To learn the $N$-to-$N$ irreflexive complementarity mapping with this embedding scheme, we use fuzzy set representations \cite{Zadeh:1965} for both embedding spaces, and encode the complementarity relation as the intersection of sets. When learning the complementary relations \emph{across} two embedding spaces, the similarity in the \emph{same} embedding space can be interpreted as interchangeability. 

\paragraph*{Key contributions:}
\begin{itemize}
\item We propose a novel dual embedding framework to learn complementarity and interchangeability relations between partial shapes.
\item The complementarity and interchangeability of partial shapes are encoded as inter- and intra-relations in the dual embedding spaces, respectively.
\item Fuzzy set representations are utilized for both embedding spaces, to learn $N$-to-$N$ irreflexive complementarity mapping between them.
\item We demonstrate the effectiveness of the proposed embedding scheme for learning the two relations between partial shapes for several shape modeling tasks, on a variety of shape categories.
\end{itemize}

\section{Related Work}
\label{sec:related_work}

We review related work on component-based 3D modeling, and structural embedding techniques using neural networks.


\paragraph*{Component-based 3D Modeling}
Funkhouser \etal~\shortcite{Funkhouser:2004} were the first to introduce the idea of reusing parts in the existing 3D models for creating new objects. Subsequent approaches~\cite{Chaudhuri:2011,Kalogerakis:2012,Chaudhuri:2013} developed the idea of shape construction using labeled components, by learning the component structure and suggesting appropriate components in the interactive assembly process. Shen \etal~\shortcite{Shen:2012} used partial scanned data as a cue for constructing complete 3D models. There, components in the database were retrieved and stitched to each other to fit to the input geometry and fill the missing area in the incomplete scans. This completion approach was further extended by Sung \etal~\shortcite{Sung:2015}, who integrated both symmetry- and retrieval-based inferences to detect the missing parts. These approaches successfully demonstrated practical applications of leveraging the component structure in a given dataset of 3D models, but all of them relied on having models with labeled components, which required significant annotation effort.

Recently, Sung \etal~\shortcite{Sung:2017} introduced a method for constructing shapes from \emph{unlabeled} components in an iterative assembly process. Specifically, given a partial shape, their method produces multiple plausible complementary components. This is done by training a neural network which jointly maps database components into the embedding space and predicts a complementary component probability distribution over that space given a partial shape. While here we also learn complementary relations, we aim at learning relations between \emph{partial shapes}, as opposed to relations between partial shapes and single components in \cite{Sung:2017}. Towards that goal, we embed all possible partial objects and discover shapes that complete the query in a plausible manner. This is more challenging problem since the shape variation space of partial shapes is much larger than the space of single components. We evaluate both methods using partial shape datasets, and demonstrate that our method outperforms the previous work in quantitative and qualitative evaluations.

\paragraph*{Deep Structural Embedding} Data embedding with neural networks is widely used for encoding relations among large-scale data. 
The advantages of inferring relations between data points in a structured embedding space, instead of learning binary indicator functions of the relations, are clearly described in the seminal work~\cite{Bordes:2011}: simple adaptation to various datasets, compact information storage, flexible joint encoding of different types of relations, and, most importantly, the ability to infer unseen relations from the structure of the embedding space. Notable examples of the above in language processing are \cite{Socher:2013} and \cite{Mikolov:2013}, where low-dimensional word embeddings were used to capture the relations among words, and in image processing - \cite{Schroff:2015}, which proposed learning an embedding space of facial images, where distances directly corresponded to face similarity, independently of the face pose.
Deep embedding of 3D data was utilized by Li \etal~\rev{\shortcite{Li:2015b}}, who suggested a method for real time object reconstruction, by learning correlation between images and 3D models, and by Wu \etal~\shortcite{Wu:2016b}, who showed that the relations among 3D object structures can be learned using a generative adversarial network. Sung \etal~\shortcite{Sung:2017} learned a different embedding space of complementary components for input partial shapes.

Unlike previous work, in the proposed approach we construct embedding spaces that reflect both complementarity relations, learned in a supervised manner, as well as interchangeability relations, for which no supervision is provided, and successfully discover both types of relations between previously unseen partial shapes.
In addition, the majority of existing techniques for deep embedding encode reflexive and symmetric relations, such as similarities, into distances or angles between vectors in the embedding space~\cite{Karpathy:2015}. Some recent methods focus on other types of relations, such partial order relations \cite{Vendrov:2016,Nickel:2017}, which are asymmetric and transitive. In this work, we consider different \emph{irreflexive} and symmetric complementarity relations, and propose a new embedding space construction technique to reflect these relations.



\section{Method}
\label{sec:method}

In this section, we explain how we train a neural network to jointly encode both the complementarity and interchangeability in embedding spaces. We first give an overview of the proposed approach, then describe in detail all its components.

\subsection{Overview}
We design binary energy functions of complementarity and interchangeability, both of which take the embedding coordinates of partial shapes as inputs. A neural network \rev{is to used to define the embedding function for partial shapes. In the training of the neural network, we create complementary pairs of partial shapes as training examples by splitting full objects into two parts. But we do not have any supervision for interchangeability. Thus, the network is trained to minimize only the complementarity energy, but is still able to predict the interchangeability from the embedding structure.}
In next subsections, we elaborate on how we define the embedding spaces and the binary energy functions on them. We first describe the motivation for using the dual embedding spaces to represent complementarity (Section \ref{sec:dual_embedding}), and how this relation is encoded across the dual spaces (Section~\ref{sec:inclusion}). Then, we define the complementarity and interchangeability energy functions as fuzzy set operations on the embedding spaces, in Sections~\ref{sec:fuzzy_set} and~\ref{sec:interchangeability}, respectively. The loss functions used for the neural network training and the neural network implementation details are described in Sections~\ref{sec:losses} and~\ref{sec:training}.

\subsection{Dual Embedding Spaces}
\label{sec:dual_embedding}
We first describe how to design the embedding spaces and a binary indicator function for complementarity.
\rev{One can consider a graph which nodes and edges indicate partial shapes and their complementary relations, respectively. Our problem can be viewed as a graph embedding problem aimed at finding unseen edges between nodes (unseen complementarity relations) by using the geometry of partial shapes as node attributes. The binary complementarity indicator function is then defined as whether a partial shape is connected to the other, and vice versa. There are several previous techniques for the graph node embedding problem with input node attributes~\cite{Perozzi:2014,Hamilton:2017,Kipf:2017}, but they all use same approach of mapping neighboring nodes to proximal locations in the embedding space, which is not applicable to our case for two reasons. First, complementarity is \emph{not} transitive, meaning that, given a partial shape, a complement of its complement is generally not a complement of it. Thus, first-order neighbors need to be discriminated from higher-order neighbors. Second, complementarity is also \emph{not} reflective; a partial shape is not a complement of itself. Thus, each node should be isolated from the its neighbors in the embedding space. To handle the \emph{non-transitivity} and \emph{irreflexivity}, we consider \emph{dual} embedding spaces as shown in Figure~\ref{fig:dual_spaces}.}
All partial shapes live in both embedding spaces, and the complementarity relations among partial shapes are now represented by the relations between their embedding coordinates \emph{across} the two spaces. Since one partial shape can have different embedding coordinates in different spaces, this allows a partial shape not to have a relation to itself.
We define the structure of two embedding spaces and the inter-space complementary relation in the next subsection.

\begin{figure*}[t!]
\centering
\includegraphics[width=0.8\linewidth]{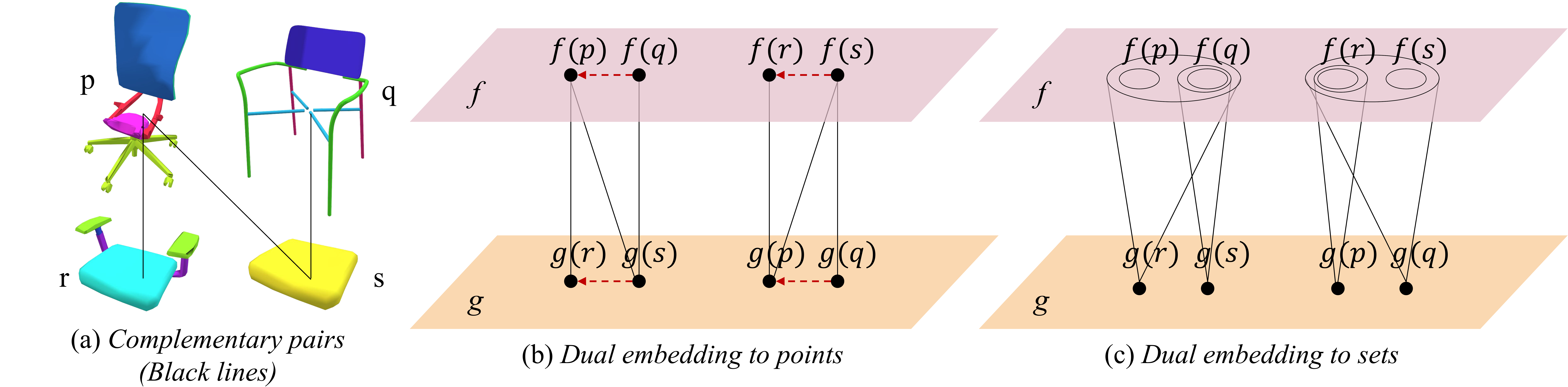} 
\caption{A schematic example showing how our set representation works on the dual embedding space. (a) (p, r) (p, s), and (q, s) are complementary pairs, and (q, r) is a non-complementary pair. (b) A case of embedding data as points, and aligning embedding coordinates of complementary shapes in different spaces (e.g. f(p) = g(s) and f(r) = g(p)).
The complementarity of $(p,r)$ and $(p,s)$ implies that $g(r) = g(s)$ (similarly, $f(r) = f(s)$). Because of the complementarity of $(q, s)$ this also \emph{incorrectly} implies that $g(q) = g(r)$ ($f(q) = f(r)$), which results in a wrong complementary relation between $q$ and $r$. 
(c) The proposed approach of representing complementarity with \emph{set representations}. All relations can be correctly encoded.}
\label{fig:inclusion}
\vspace{-0.5cm}
\end{figure*}

\subsection{Embedding as Set Inclusion}
\label{sec:inclusion}
The na\"ive idea for constructing dual spaces is to align the positions of complementary partial shapes at the same coordinates in different spaces.  However,
\rev{it may cause some non-complementary pairs to be encoded as complementary to each other.}
For example, in the case illustrated in Figure~\ref{fig:inclusion}(a), shape $p$ is complementary to both $r$ and $s$, but shape $q$ is only complementary to $s$. Then, from the complementarity of $p$ and $r,s$, both $r$ and $s$ should be placed at the location of $p$ in the other space ($f(p) = g(r) = g(s)$ and $f(r) = f(s) = g(p)$). Because of the complementarity of $q$ and $s$, $q$ also goes to $s$'s location in the other space. This leads to a complementary relation between $q$ and $r$, which is contradictory to the assumption (See Figure~\ref{fig:inclusion}(b)).

To avoid this problem, we suggest to relate a coordinate in one embedding space to \emph{multiple} coordinates in the other space.
\rev{In the aspect of graph embedding described in Section~\ref{sec:dual_embedding}, the relation is indicated by checking both ways of whether a node is a neighbor of the other node. When considering one-way relations only, and encoding $1$-to-$N$ mapping from a partial shape to its complements,} we view the embedding coordinates as a representation of a \emph{set}, such that the complementarity relation is encoded as \emph{inclusion} from one space to the other.
One choice of encoding sets and inclusions in the embedding space is using the approach of Vendrov \etal~\shortcite{Vendrov:2016}:

\vspace{-0.5cm}
\begin{align}
x \subseteq y \quad \Leftrightarrow \quad \bigwedge_{i=1}^{D} x_i \leq \rev{y_i},
\label{fig:inclusion_embeddding}
\end{align}

where $x, y \in \mathbb{R}^N_{+}$ are the embedding coordinates of $x,y$, respectively\rev{, and $\bigwedge$ is a `logical and' operator} (note that Vendrov \etal \rev{use} reversed direction of the inequality but we switch back to the natural direction). Since here we wish to relate embedding coordinates in two different spaces (due to the irreflexivity), given two embedding spaces $f(\cdot)$ and $g(\cdot)$ we represent the
\rev{\emph{one-way}} complementarity $c(x \rightarrow y)$ as follows:  

\vspace{-0.5cm}
\begin{align}
c(x \rightarrow y) \quad \Leftrightarrow \quad \bigwedge_{i=1}^{D} f(x)_i \leq g(y)_i .
\label{eq:dic_comp_rep}
\end{align}

According to the analogy with the inclusion relation, we will call $f$ and $g$ the \emph{subset} and \emph{superset} spaces, respectively, in the rest of paper. Then, the binary indicator function for
\rev{\emph{both-way}} complementarity $c(x, y)$ can be represented
\rev{as follows}:

\vspace{-0.3cm}
\begin{align}
c(x, y) \quad \Leftrightarrow \quad c(x \rightarrow y) \wedge c(y \rightarrow x) .
\label{eq:undic_comp_rep}
\end{align}

Figure~\ref{fig:inclusion}(c) illustrates how the relations between $p,q,r,s$ in Figure~\ref{fig:inclusion}(a) can be represented with sets and inclusions.


\subsection{Fuzzy Set Interpretation}
\label{sec:fuzzy_set}
While the inclusion embedding of \cite{Vendrov:2016} was first applied for neural network training in that paper, the idea is actually closely related to the well-studied \emph{fuzzy set theory} \cite{Zadeh:1965}. In fuzzy set theory, a set is represented with fuzzy memberships over a discrete set of elements, which is also called \emph{possibility} distribution. Then, the notion of inclusion is defined so that the membership scores (or probabilities) over all elements of a superset are greater or equal to the corresponding scores of a subset, which is identical with the definition in Equation~\ref{fig:inclusion_embeddding}.

With the fuzzy set representation, one can consider how to define set operations analogous to classic crisp (non-fuzzy) set theory. For example, for `logical and' (intersection) $\wedge$, and `inclusive or' (union) $\vee$, there are various ways of defining the operations with fuzzy sets (see Section~3 in \cite{Zimmermann:2001} for details), but the simplest form is using minimum and maximum operations:

\vspace{-0.3cm}
\begin{equation}
\begin{aligned}
(x \wedge y)_i &= \min(x_i, y_i) , \\
(x \vee y)_i &= \max(x_i, y_i) .
\end{aligned}
\end{equation}

In the neural net training, we need to \emph{fuzzify} the notion of inclusion to obtain a continuous loss function. Vendrov \etal suggest to penalize when the embedding coordinates of a subset are greater than the coordinates of the superset, element-wise:

\vspace{-0.3cm}
\begin{align}
E(x \subseteq y) = \sum_i \max(0, x_i - y_i)^2 .
\end{align}

This is actually the same as making the subset to be equal to the intersection of the two sets $x$ and $y$, using the above definition of intersection:

\vspace{-0.3cm}
\begin{equation}
E(x \subseteq y) = \sum_i \left(x_i - \min(x_i, y_i)\right)^2 = \| x - (x \wedge y) \|_2^2 .
\end{equation}

Using Equation~\ref{eq:dic_comp_rep} and Equation~\ref{eq:undic_comp_rep}, we can define the energy functions for the \rev{one-way} complementarity $E_c(x \rightarrow y)$ and the \rev{both-way} complementarity $E_c(x, y)$, as follows:

\vspace{-0.3cm}
\begin{equation}
E_c(x \rightarrow y) = \sum_i \max(0, f(x)_i - g(y)_i)^2 ,
\label{eq:come_energy}
\end{equation}

\vspace{-0.3cm}
\begin{equation}
E_c(x, y) = E_c(x \rightarrow y) + E_c(y \rightarrow x) .
\label{eq:undic_come_energy}
\end{equation}

\subsection{Interchangeability via Complementarity}
\label{sec:interchangeability}
In Section \ref{sec:inclusion} and \ref{sec:fuzzy_set}, we described how complementarity is represented as an \emph{inter}-space fuzzy set operation. Now we discuss how we define an \emph{intra}-space fuzzy set operation that measures the degree of \emph{interchangeability}.

It is obvious that two partial shapes have exactly same complements with same energy values when they have the same embedding coordinates in both embedding spaces. This implies that two partial shapes with similar embedding coordinates in each embedding space have a similar set of complements.
In the following propositions, we show that how the \emph{union} and \emph{intersection} of two fuzzy sets represented by the embedding coordinates are related to the complementarity energies with arbitrary partial shapes.

\begin{proposition}
$\max \left( E_c(x \rightarrow z), E_c(y \rightarrow z) \right) \leq E_c((x \vee y) \rightarrow z)$
\label{prob:bound}
\end{proposition}
\begin{proof} Refer to Appendix
\end{proof}

\begin{proposition}
$\max \left( E_c(z \rightarrow x), E_c(z \rightarrow y) \right) \leq E_c(z \rightarrow (x \wedge y))$
\end{proposition}
\begin{proof} Refer to Appendix
\end{proof}

\begin{corollary}
\begin{align*}
&\frac{1}{2}\left(E_c(x, z) + E_c(y,z)\right) \\
&\leq \max \left( E_c(x \rightarrow z), E_c(y \rightarrow z) \right) + \max \left( E_c(z \rightarrow x), E_c(z \rightarrow y) \right) \\
&\leq E_c((x \vee y) \rightarrow z) + E_c(z \rightarrow (x \wedge y))
\end{align*}
\label{cor:upper_bound}
\end{corollary}

Corollary~\ref{cor:upper_bound} shows that the union on the $subset$ space and the intersection on the $superset$ space bound the sum of complementarity energies for arbitrary partial shapes.
When restrict the $l2$-norm of all embedding coordinates to be one, one can consider how close the $l2$-norm of $f(x) \vee f(y)$ and $g(x) \wedge g(y)$ are to one as a measure of \emph{interchangeability} energy $E_r(x,y)$:

\vspace{-0.3cm}
\begin{equation}
\begin{aligned}
E_r(x,y) &\triangleq (\|f(x) \vee f(y) \|^2_2 - 1) + (1 - \|g(x) \wedge g(y) \|^2_2) \\
&= \|f(x) \vee f(y) \|^2_2 - \|g(x) \wedge g(y) \|^2_2
\end{aligned}
\label{eq:interchangeability_measure}
\end{equation}

Note that the unit $l2$-norm constraint is not only for defining interchangeability relation in the embedding, but it is also a common constraint in neural network training to avoid over-fitting \cite{Schroff:2015,Vendrov:2016}.

\subsection{Neural Network Loss}
\label{sec:losses}
The loss function for the neural network training \rev{is defined using} the complementarity energy (Equation~\ref{eq:undic_come_energy}).
Given $N$ complementary pairs ${(x_i, y_i)}$ in a batch, we consider all mis-matched pairs $(x_i, y_{j \neq i})$ as negative examples.
We suggest two loss functions that can be used depending on the application. For complement retrieval tasks, we use pairwise ranking loss as introduced in \cite{Karpathy:2015,Vendrov:2016}:

\vspace{-0.3cm}
\begin{equation}
\begin{aligned}
L_R &= \sum_i \sum_{j \neq i} \max\left(0, E_c(x_i, y_i) - E_c(x_i, y_j) + \alpha \right) \\
&+ \sum_j \sum_{i \neq j} \max\left(0, E_c(x_i, y_i) - E_c(x_i, y_j) + \alpha \right), \\
\end{aligned}
\label{eq:ret_loss}
\end{equation}

where $\alpha$ is a given margin parameter. We use $\alpha=0.05$ in all our experiments.
This pairwise ranking loss learns \emph{relative} distances between positive and negative pairs. But we need a commensurable measure of complementarity for the interchangeability measure (Equation~\ref{eq:interchangeability_measure}) since it is based on the upper bound of the complementarity energies. Thus, we introduce another loss function learning absolute errors with a threshold:

\vspace{-0.3cm}
\begin{equation}
\begin{aligned}
L_T &= \frac{1}{N} \sum_i \max\left(0, E_c(x_i, y_i) - \left(t - \frac{1}{2}\alpha\right)\right) \\
&+ \frac{1}{N(N-1)} \sum_i \sum_{j \neq i} \max\left(0, \left(t + \frac{1}{2}\alpha\right) - E_c(x_i, y_j)\right),
\end{aligned}
\label{eq:thr_loss}
\end{equation}

where $t$ is a learnable threshold parameter. This loss function makes the energy of positive pairs smaller than $t$ and the energy of negative pair greater than $t$ with the margin $\alpha$ between them.



\subsection{Neural Network Architecture and Training}
\label{sec:training}
\begin{figure}[t!]
\centering
\includegraphics[width=\linewidth]{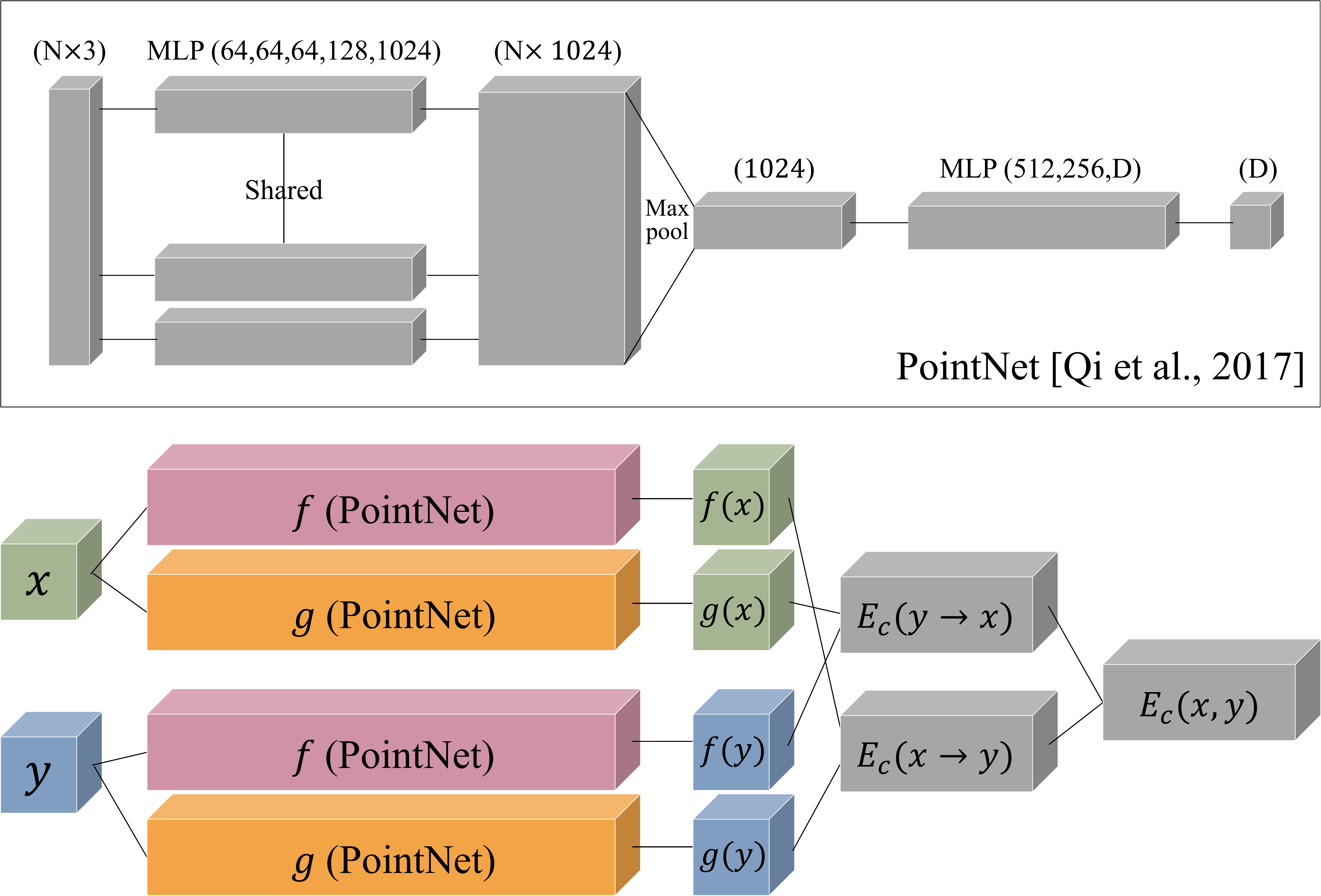}
\caption{Neural network architecture. We build two Siamese networks with shared weights, for both $f$ and $g$ embeddings. We use PointNet~\shortcite{Qi:2017} to compute the embeddings, but any neural network architecture for 3D geometry processing is applicable.}
\label{fig:network}
\vspace{-0.5cm}
\end{figure}

We used the PointNet architecture~\shortcite{Qi:2017} as a basic building block of the proposed embedding network, as illustrated in Figure~\ref{fig:network}. Since all partial shapes are embedded into both embedding spaces, we have two separate PointNet Siamese architectures for learning the $f$ and $g$ embedding functions.
We feed both $f$ and $g$-networks with $1k$ sampled points $x$ and $y$, to produce $f(x), g(x)$ and $f(y), g(y)$. Note that we use the unit $l2$-norm constraint for output embedding coordinates as described in Section \ref{sec:interchangeability}. Both networks receive inputs as \emph{centered} point clouds of partials shapes, translated so that the centers of axis-aligned bounding boxes are located at the origin. Thus, the relation prediction is performed without the information of the partial shape location.
In all experiments, we train the network for 1,000 epochs with batch size of 32, and use ADAM optimizer with 1.0E-3 initial learning rate, 0.7 decay rate and 200K decay steps. The embedding dimension is fixed to be $D = 100$.

\section{Results}
\label{sec:results}

\subsection{Partial Shape Dataset}
In our experiments, we use the ShapeNet 3D component dataset of Sung \etal \shortcite{Sung:2017}. It consists of 9 model categories from the ShapeNet \cite{Chang:2015}, \rev{each of which} has up to 2.4K models. \rev{All models are consistently aligned, scaled to have unit radius, and pre-segmented into components.} The components were created from the ShapeNet CAD model scene graphs; the leaf node components of scene graphs were preprocessed, so that symmetric components were grouped into a single component, and small components were merged with the adjacent larger components. We build \emph{contact} graphs of components based on their proximity, and during training generate complementary partial shape pairs by randomly splitting the contact graphs into two subgraphs. The dataset is split into $80\%$ training and $20\%$ test sets, and separate networks are learned for different model categories.

\begin{figure*}[t!]
\centering
\includegraphics[width=\linewidth]{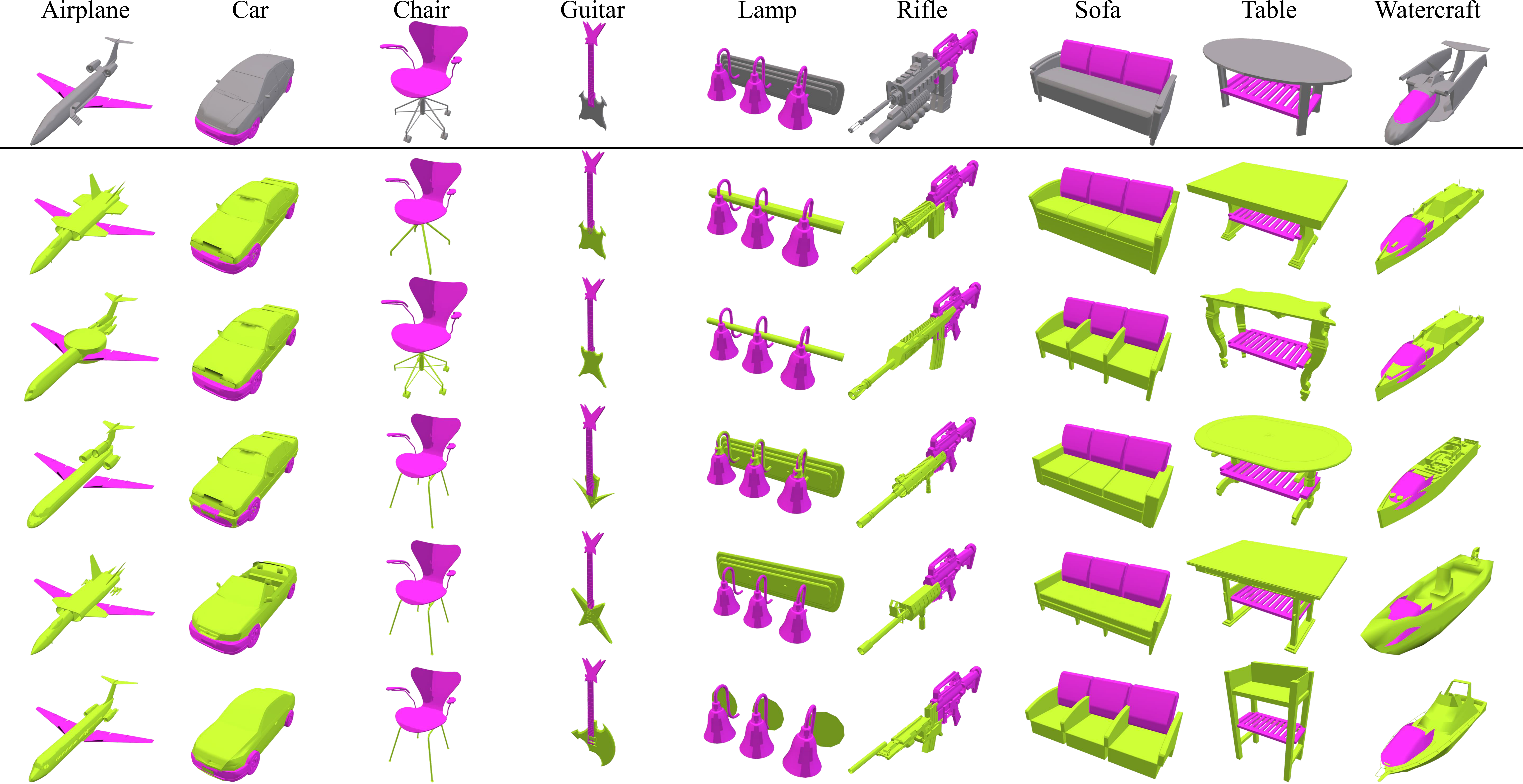} 
\caption{Examples of top-5 complement shape retrievals. The top row shows the original shapes (gray) with highlighted query partial shapes  (magenta). Next five rows show the retrieved top-5 complementary partial shapes (green), together with the query shapes (magenta). See the accompanying text for details.}
\label{fig:complementarity}
\end{figure*}

\begin{figure*}[t!]
\centering
\includegraphics[width=\linewidth]{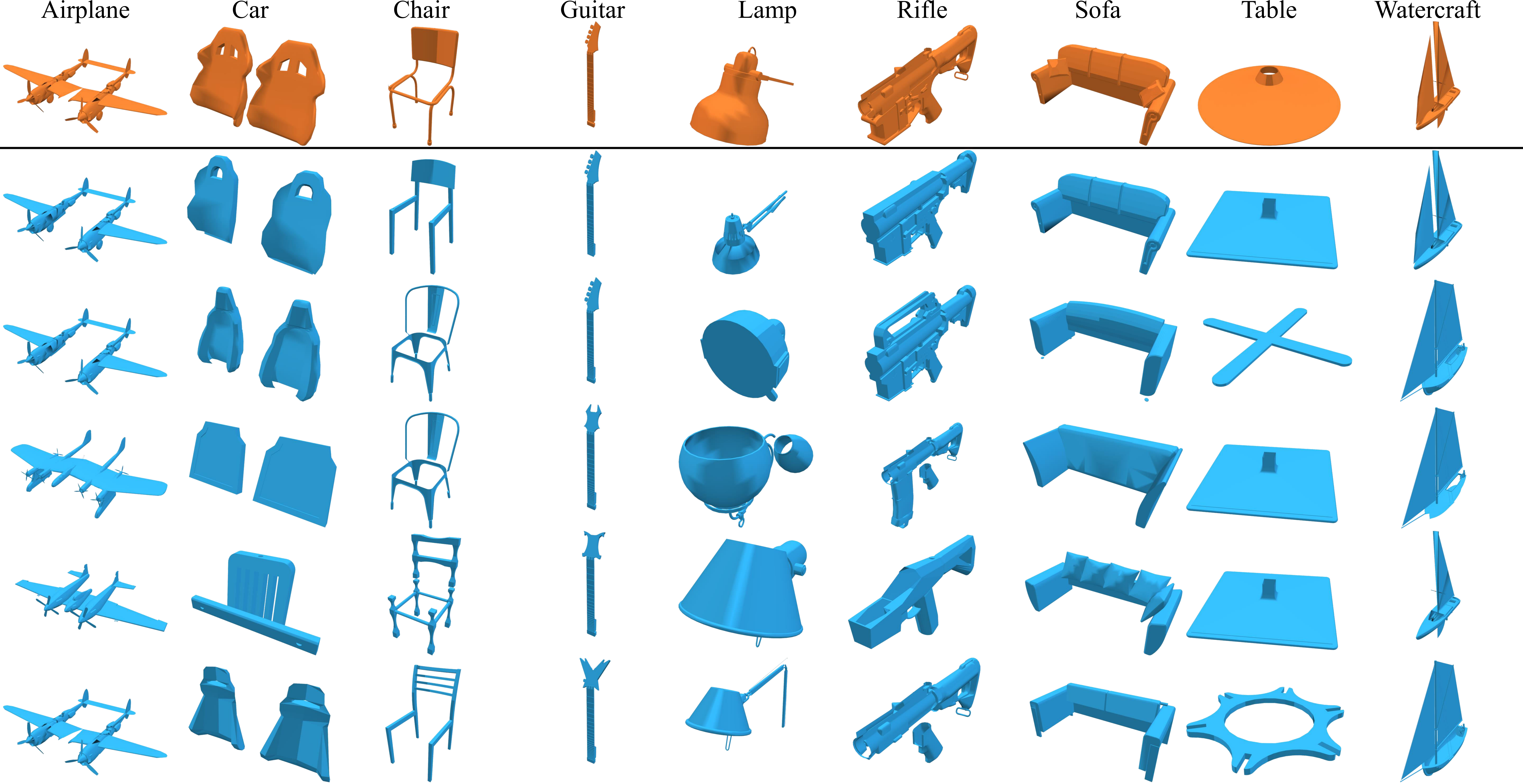} 
\caption{Examples of partial shape interchangeability. The top row shows the query partial shapes. Next five rows show partial shapes interchangeable with it, which were detected using the proposed approach. Different components comprising the partial shapes are shown with different shades of blue. See the accompanying text for details.}
\label{fig:interchangeability}
\end{figure*}

\subsection{Qualitative Evaluation}
\label{sec:qualitative}
\paragraph*{Complementarity Evaluation}
Figure~\ref{fig:complementarity} shows examples of the top-5 complement retrieval, in terms of
\rev{the} complementarity energy $E_c$~(Equation~\ref{eq:undic_come_energy}), one example per category. In the retrieval experiment, we used all possible partial shapes in the test set as a database of complement candidates. The centered query and retrieved partial shapes were automatically stitched, using the placement neural network introduced in \cite{Sung:2017}. Here, the placement net was re-trained with partial shapes instead of single components. We note that the retrieved complementary partial shapes have different geometries and styles, but most of them complement the queries in a plausible manner. For example, given a chair query with swivel legs removed, both four legs and swivel legs are retrieved, and all the results look plausible. In \rev{another} example, given the stretcher of a table, different tables \rev{having} appropriate widths fitting the stretcher are retrieved. The three lamps are complemented with suitable mount accessories, including the three wall mount plates of the last complement.

\paragraph*{Interchangeability Evaluation} Figure~\ref{fig:interchangeability} shows examples of interchangeable partial shapes extraction. Here, we also used all possible partial shapes in the test set as a database of interchangeable candidates. For each query shape, we extracted its top-5 nearest partial shape neighbors, now using the interchangeability measure $E_r$ (Equation~\ref{eq:interchangeability_measure}).
The results demonstrate that our method can properly learn interchangeability among partial shapes, even when they are constructed of different components and have dissimilar geometries. For instance, table stand bottoms (Figure~\ref{fig:interchangeability}, second column from the right) have different shapes,  but they can be replaced by each other in any table. For the partial chair without a seat, we successfully retrieve all partial chairs without seats, while all retrievals have different back and leg parts. The lamp components retrieved by a query have various shapes with different sizes, but all of them are shade parts with tubings.

\begin{table*}[t!]
\centering
\begin{tabular}{c|c|*{9}c|c}
\toprule
  \multicolumn{2}{c|}{\shortstack{Category\\\small{(\# Partial Shapes)}}}  &  \shortstack{Airplane\\(4140)}  &  \shortstack{Car\\(5770)}  &  \shortstack{Chair\\(8374)}  &  \shortstack{Guitar\\(198)}  &  \shortstack{Lamp\\(1778)}  &  \shortstack{Rifle\\(1184)}  &  \shortstack{Sofa\\(4452)}  &  \shortstack{Table\\(4594)}  &  \shortstack{Watercraft\\(1028)}  & Mean  \\
\midrule
  \multirow{2}{*}{Recall@1} & CM  & 9.9 & 2.4 & 4.9 & 19.2 & 1.7 & 1.9 & 3.9 & 2.7 & 0.7 & 4.3\\
  & Ours &  \textbf{17.5}  &  \textbf{5.8}  &  \textbf{8.0}  &  \textbf{23.7}  &  \textbf{5.1}  &  \textbf{7.3}  &  \textbf{6.7}  &  \textbf{4.1}  &  \textbf{3.2}  &  \textbf{7.8} \\
\midrule
  \multirow{2}{*}{Recall@10} & CM  & 48.6 & 15.5 & 27.2 & 67.7 & 11.1 & 17.1 & 20.0 & 15.5 & 7.3 & 23.5\\
  & Ours &  \textbf{61.3}  &  \textbf{30.5}  &  \textbf{35.0}  &  \textbf{72.2}  &  \textbf{19.7}  &  \textbf{23.5}  &  \textbf{30.1}  &  \textbf{19.2}  &  \textbf{14.3}  &  \textbf{32.9} \\
\midrule
  \multirow{2}{*}{\shortstack{\small{Median}\\\small{Percentile Rank}}} & CM  & 99.8 & 98.8 & 99.6 & 97.0 & 89.6 & 94.3 & 98.5 & 98.3 & 87.0 & 97.9\\
  & Ours &  \textbf{99.9}  &  \textbf{99.5}  &  \textbf{99.7}  &  \textbf{98.5}  &  \textbf{90.4}  &  \textbf{95.8}  &  \textbf{99.2}  &  \textbf{98.5}  &  \textbf{88.7}  &  \textbf{98.4} \\
\midrule
  \multirow{2}{*}{\shortstack{\small{Mean}\\\small{Percentile Rank}}} & CM  & 98.4 & 96.4 & 98.3 & \textbf{94.5} & \textbf{81.4} & 88.2 & 94.0 & 94.9 & 77.6 & 94.8 \\
  & Ours &  \textbf{98.5}  &  \textbf{97.2}  &  \textbf{98.5}  &  93.8  &  79.9  &  \textbf{89.0}  &  \textbf{94.9}  &  \textbf{95.0}  &  \textbf{78.7}  &  \textbf{95.2} \\
\bottomrule
\end{tabular}
\caption{Quantitative evaluations of complement retrievals using ComplementMe~\shortcite{Sung:2017} and our method. The number of all possible partial shapes in the test set are given in parentheses at the first row. 
Recall@$N$ measures the percentage of the ground truth complements in the top-$N$ rank retrievals, and Percentile Rank measures the percentage of partial shapes having ranks equal or greater than the rank of the ground truth complements. Higher is better in all measures.}
\label{tbl:complementarity_comparisons}
\end{table*}

\begin{figure*}[t!]
\centering
\includegraphics[width=\linewidth]{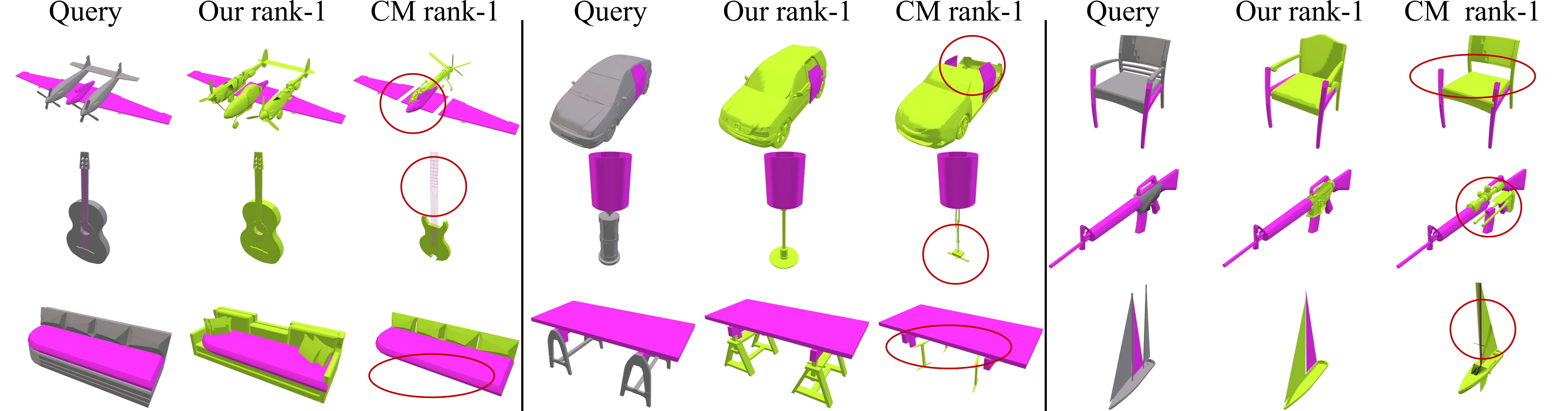} 
\caption{Comparison of complement shape retrieval results of the proposed method and the ComplementMe architecture.
In each case, the first column is the query partial shape, the second column is our rank-1 retrieval, and the third column is the rank-1 retrieval of ComplementMe~\shortcite{Sung:2017}. The color-coding is the same as in Figure~\ref{fig:complementarity}. The red circles show incomplete areas in the ComplementMe results.
}
\label{fig:complementarity_comparisons}
\vspace{-5mm}
\end{figure*}

\begin{figure}[t!]
\centering
\includegraphics[width=\linewidth]{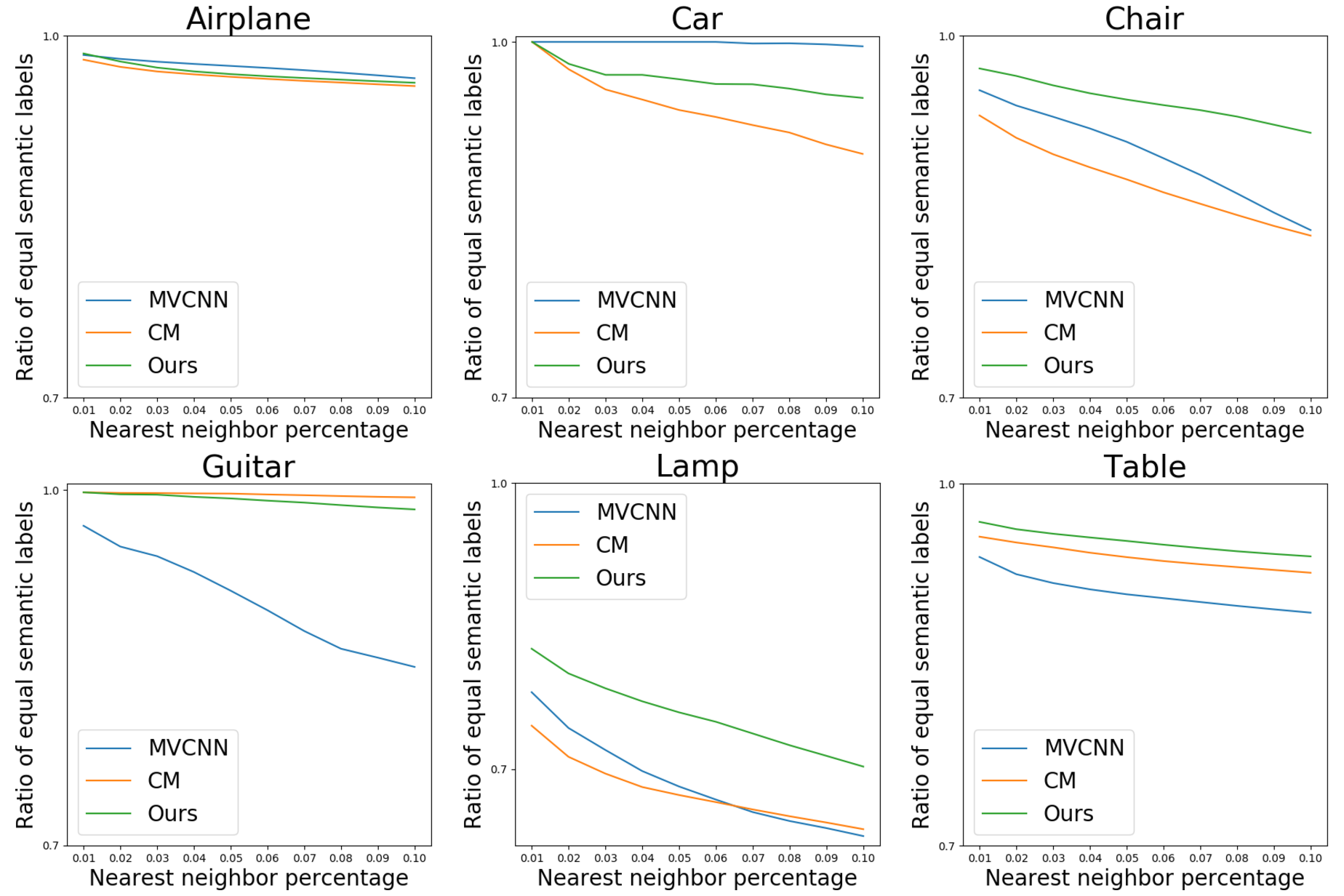} 
\caption{Evaluation of interchangeability measures with human-annotated part correspondences in \cite{Yi:2016}. We present correlations between the interchangeability measures and the semantic part labels. Our method (green) shows better performance, indicated by higher correlations, as compared to MVCNN (blue)~\cite{Su:2015} and ComplementMe (orange)~\shortcite{Sung:2017}. See the accompanying text for details.}
\label{fig:interchangeability_comparisons_graph}
\vspace{-5mm}
\end{figure}

\begin{figure*}[t!]
\centering
\includegraphics[width=\linewidth]{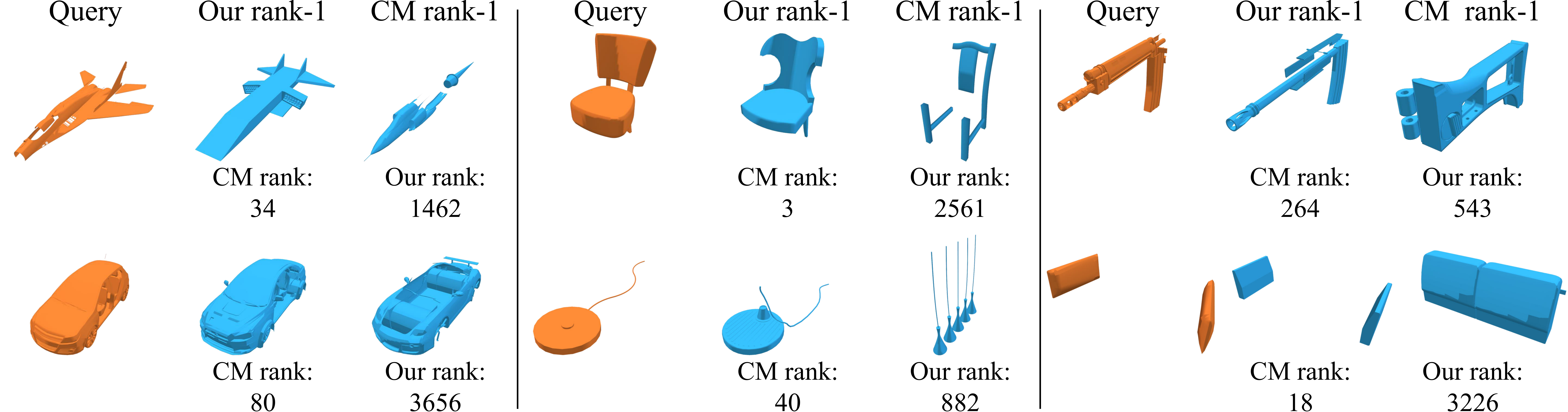} 
\caption{Comparison of rank-1 nearest neighbors detected using our interchangeability measure, and distances in the embedding space of ComplementMe~\shortcite{Sung:2017}. The numbers below show the ranks when using different methods. ComplementMe sometimes finds wrong interchangeable partials shapes while ours correctly discriminates correct and wrong interchangeable shapes.}
\label{fig:interchangeability}
\vspace{-0.2cm}
\end{figure*}

\begin{figure*}[t!]
\centering
\includegraphics[width=\linewidth]{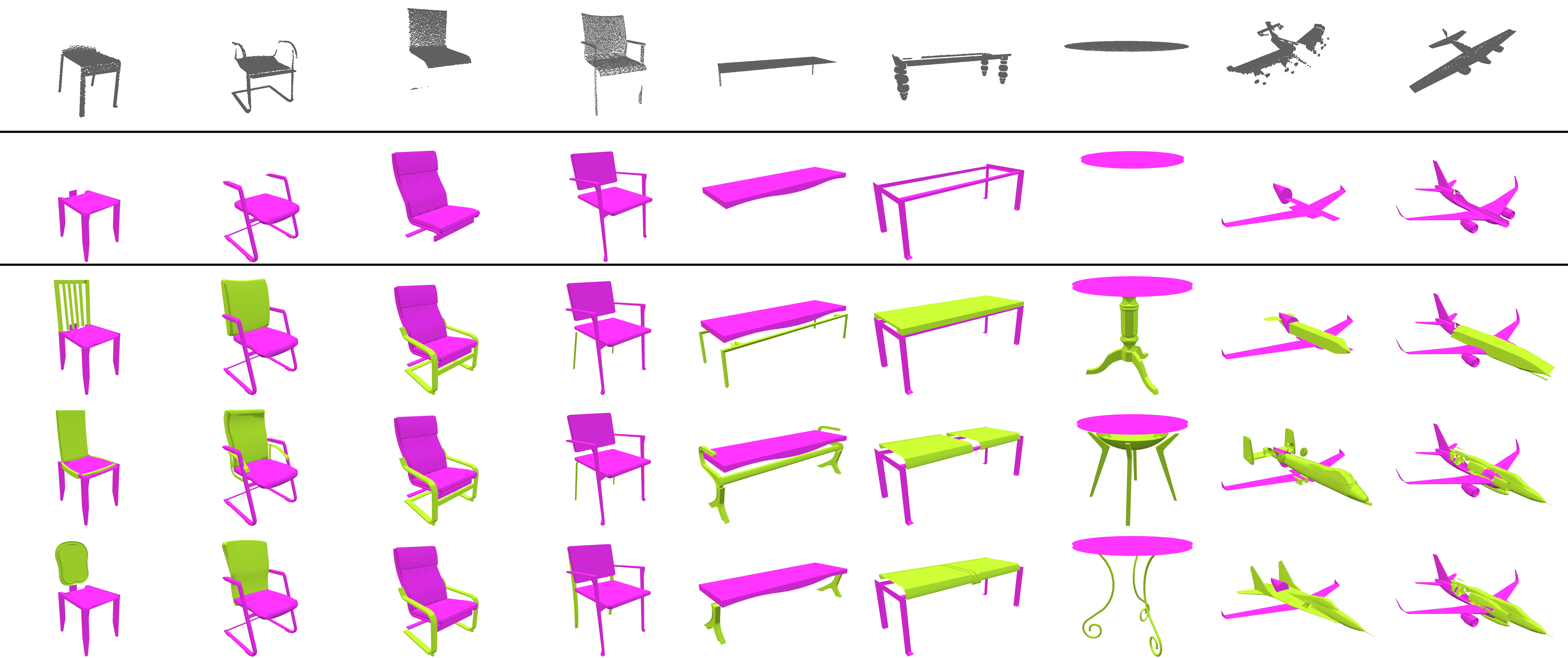} 
\caption{Examples of partial scan completion. We used synthetic partial scan data from the benchmark of \cite{Sung:2015} (first row). We first use ICP to retrieve the best fitting partial shape from our database (second row, colored in pink), and then complete it to a full shape using various complements retrieved by the proposed method (colored in green). Three out of the best ten completions are shown in third, fourth, and fifth rows. The positions of the complements are automatically predicted using the placement network of~\rev{\cite{Sung:2017}}.}
\label{fig:completion}
\vspace{-0.5cm}
\end{figure*}

\subsection{Comparisons}
We compare our method with ComplementMe \shortcite{Sung:2017}, which also learns complementary relations among 3D shapes, but for a different purpose.
ComplementMe was designed for an iterative component assembly task, therefore it retrieves a single \emph{component} for each query partial shape
\rev{in every iteration. Thus, when applied to fully automatic shape synthesis or completion, ComplementMe has limitations of accumulating noise in successive iterations and missing a notion of termination. Contrarily, the proposed method finds \emph{groups of components} fully completing the query shape in a single retrieval step. Another difference in terms of the difficulty of the problems is that} the proposed method handles much larger shape variation space since it embeds all possible partial shapes, while ComplementMe only embeds single components.

\rev{ComplementMe approach can also be adapted to handle partial shapes instead of individual components. However, we argue that the proposed method is more effective for learning both complementarity and interchangeability relations due to the differences in relation representations. In ComplementMe, the one-way complementarity energy function is defined as a negative log-likelihood of a Gaussian mixture probability density function. The multi-modality of the distribution is essential in ComplementMe since a single Gaussian raises the embedding collapse problem described in Section~\ref{sec:dual_embedding} and Figure~\ref{fig:inclusion}(b). But it also leads to two limitations. First, a larger number of Gaussians can better encode all possible complementary relations, but it also increases the number of output parameters, making the network more difficult to train. Thus, the representation power can be impaired either when the number of Gaussians is too small or too large. Second, some interchangeability relations may not be captured with the multi-modal distributions since two interchangeable partial shapes can be included in different modes. Our fuzzy-set-based representation, using a single vector to represent a partial shape, is much more concise than the Gaussian mixture representation and does not have the above multi-modality issues. }

\rev{In the following experiments, we demonstrate that the difference in the relation representations affects the performance of both complementary and interchangeable shape retrievals in practice.}
For comparison, we re-train the ComplementMe retrieval network using our partial shape dataset, and the same training parameters and embedding dimension as in the proposed method. We also use our loss functions (Equation~\ref{eq:ret_loss} and Equation~\ref{eq:thr_loss}) instead of the triplet loss in ComplementMe. Note that the losses are identical except for the larger number of negative pairs used in the proposed loss ($N(N-1)$ vs. $N$ in ComplementMe, which makes the training more efficient).

\rev{Evaluating whether the retrieved shapes are complementary or interchangeable is non-trivial since the criteria are subjective. Human annotations may not be consistent and can be prone to bias. Thus, for the quantitative evaluation of the complement retrievals, we measure recall and rank of the \emph{ground truth} complements, following the recent retrieval work~\shortcite{Klein:2015,Vendrov:2016}.
Table \ref{tbl:complementarity_comparisons} shows the quantitative evaluation results when testing both ComplementMe and our method with all possible partial shapes in the test set as queries} (the partial shape number are in parentheses at the first row, for each shape category).
\rev{Recall@$N$ indicates the percentage of the ground truth in the top rank-$N$ retrievals, and percentile rank indicates the percentage of partial shapes having ranks equal or greater than the rank of the ground truth.} 
The proposed fuzzy set representation outperforms ComplementMe in all cases, except for mean percentile ranks for two model categories.

\rev{Figure~\ref{fig:complementarity_comparisons} shows complementary shape retrieval results of both methods.} Although ComplementMe produces reasonable results, some retrievals do not fully complement the query shape, resulting in missing areas in the combined shapes: e.g., armrests in a chair, a trunk in a car, and the bottom parts of a sofa and a lamp. Our method successfully creates complete plausible output shapes with the retrieved complements.

\rev{For the interchangeable shape retrievals, we also compare the proposed method with ComplementMe, and additionally with Multi-View CNN (MVCNN) descriptor \cite{Su:2015}. To evaluate the retrievals quantitatively, we use \emph{semantic} single parts \cite{Yi:2016} instead of partial shapes, and measure the correlation between the detected interchangeable parts and their semantic labels. While this measure is imperfect since some parts with different labels may have similar shapes, this is uncommon for most shape categories.
Given a query semantic part, we find its interchangeable parts using each one of the methods: the interchangeability measure $E_r(x, y)$ (Equation~\ref{eq:interchangeability_measure}) for our method, and the distances in the embedding and descriptor spaces for ComplementMe and MVCNN, respectively. Then, we compute the ratio of the number of neighbors having the same semantic label as the query and the number of all retrieved neighbors.}
Higher values mean that more neighboring parts share the semantic label with the query. Figure~\ref{fig:interchangeability_comparisons_graph} shows the average ratio of equal nearest neighbor labels, as a function of the number of the neighbors, up to 10\% of the all parts. Our method (green line) consistently produces higher correlation with semantic labels compared to other methods in all categories, except cars, for which training data mostly have coarse segmentations not into parts but into larger partial shapes.


Figure~\ref{fig:interchangeability} visualizes some results of interchangeable partial shape retrievals using both methods: query shape, rank-1 retrieval of ours, and rank-1 retrieval of ComplementMe, from left to right.
We also list the retrieval ranks based on the interchangeability measure (embedding distances for ComplementMe and energy function $E_r(x, y)$ for ours) of the other method.
The results indicate that the proposed method is able to detect more semantically meaningful shapes as its rank-1 retrievals, as compared to ComplementMe. Furthermore, according to the ComplementMe ranks, it maps both the shapes incorrectly retrieved by it and the plausible shapes retrieved by the proposed method nearby in its embedding space. In contrast, the proposed method is able to discriminate between the interchangeable and non-interchangeable shapes, as indicated by its ranks for the shapes retrieved by ComplementMe.

\subsection{Application - Partial scan completion}
One potential application of the proposed method is completion of partial scan data with various partial shapes form the dataset. Figures~\ref{fig:teaser} and~\ref{fig:completion} show examples of completing real and synthetic partial scan data with complements retrieved with our method. 
We first use ICP with manual initial scan pose, to find a partial shape in our test dataset that best fits the input point cloud (shown with pink). We then retrieve complementary partial shapes (shown with green) using our fuzzy set operations. Note that, unlike other shape completion methods, such as \cite{Sung:2015,Dai:2017}, which create a single completion result, our method can easily provide multiple plausible outputs according to the partiality of the input data.

\subsection{Computation time}
Both training and test are performed with a single NVIDIA TITAN Xp graph card. It took 4.5 hours to train a network for 50$k$ iterations. At test time, computing the embedding coordinates of 5$k$ partial shapes and their corresponding complementarity/interchangeability energies w.r.t. the query, takes in a few seconds.




\section{Conclusion and future work}
\label{sec:conclusion}
We have presented a novel neural network-based framework for learning complementarity and interchangeability relations between partial shapes.
The two relations are learned jointly by embedding the partial shapes into dual embedding spaces, where the shapes are encoded using the fuzzy set representation.
\rev{This embedding} allows us to model the complementarity and the interchangeability as fuzzy set operations performed across and within the embedding spaces, respectively. The method is fully automatic, and was trained using a dataset of models with unlabeled components. Qualitative and quantitative evaluations demonstrate that our method captures well both types of relations, and produces meaningful results when applied to previously unseen shapes.

\begin{figure}[t!]
\centering
\includegraphics[width=\linewidth]{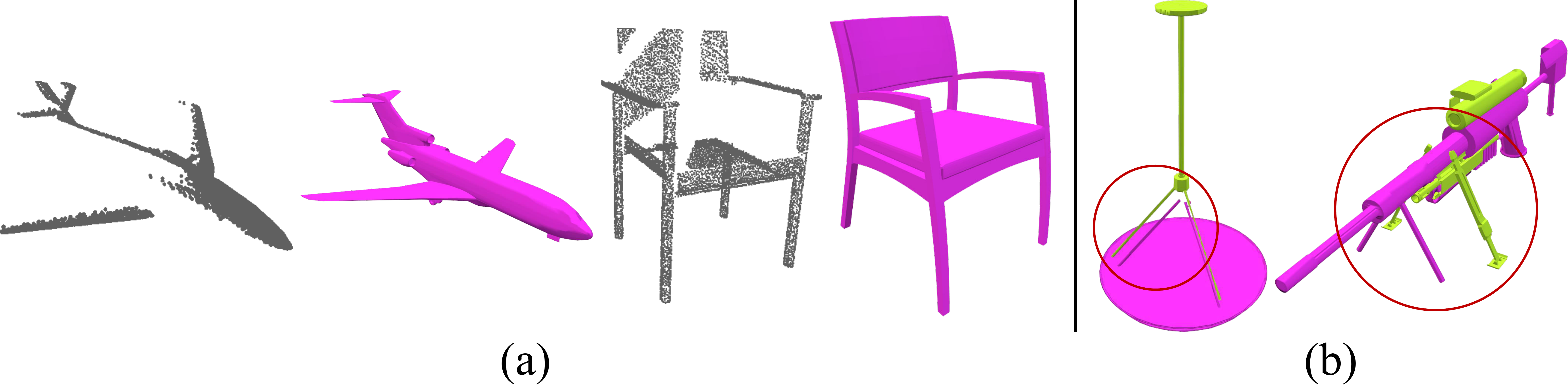} 
\caption{\rev{Limitations of our method. (a) Our complement retrieval is not applicable to complete scans when they include all components but have small missing areas. Left is a scan, and right is a ICP retrieval result as done in Figure~\ref{fig:completion}. (b) Small or thin parts may not be recognized well due to the limited resolution of point clouds, and thus the retrieved complements may include duplicate parts with the query (circled in red).}}
\label{fig:limitation}
\vspace{-0.7cm}
\end{figure}

\rev{While our framework is applicable to partial shape completion, it is limited to filling the missing area at the level of the components, and does not facilitate symmetry information as done in~\cite{Sung:2015} (Figure~\ref{fig:limitation}(a)). Also, small or thin components can be neglected in the retrieved shapes due to the limited resolution of point clouds used as neural network inputs (Figure~\ref{fig:limitation}(b)). }

In future work, we plan to investigate how the fuzzy set operations can be applied to represent the other shape relations, and also the relations among different modalities, e.g. images and 3D shapes.

\rev{
\paragraph*{Acknowledgements}
We thank the anonymous reviewers for their comments and suggestions. M. Sung acknowledges the support in part by the Korea Foundation for Advanced Studies. A. Dubrovina acknowledges the support in part by The Eric and Wendy Schmidt Postdoctoral Grant for Women in Mathematical and Computing Sciences. L. Guibas acknowledges the support of NSF grants IIS-1528025, DMS-1521608, and DMS-1546206, and gifts from the Adobe systems, Amazon AWS and Autodesk corporations.
}

\bibliographystyle{eg-alpha-doi}

\bibliography{paper}


\section*{Appendix}
\label{sec:appendix}

\paragraph*{Proposition 1}
$\max \left( E_c(x \rightarrow z), E_c(y \rightarrow z) \right) \leq E_c((x \vee y) \rightarrow z)$
\label{prob:bound}

\begin{proof}
\begin{align*}
&\max \left( E_c(x \rightarrow z), E_c(y \rightarrow z) \right) \\
&= \max \left( \sum_i \max(0, f(x)_i - g(z)_i)^2, \sum_i \max(0, f(y)_i - g(z)_i)^2 \right) \\
&\leq \sum_i \max \left( \max(0, f(x)_i - g(z)_i)^2, \max(0, f(y)_i - g(z)_i)^2 \right) \\
&= \sum_i \max \left( 0, \max(f(x)_i, f(y)_i) - g(z)_i \right)^2 \\
&= E_c((x \vee y) \rightarrow z)
\end{align*}
\end{proof}

\paragraph*{Proposition 2}
$\max \left( E_c(z \rightarrow x), E_c(z \rightarrow y) \right) \leq E_c(z \rightarrow (x \wedge y))$

\begin{proof}
\begin{align*}
&\max \left( E_c(z \rightarrow x), E_c(z \rightarrow y) \right) \\
&= \max \left( \sum_i \max(0, f(z)_i - g(x)_i)^2, \sum_i \max(0, f(z)_i - g(y)_i)^2 \right) \\
&\leq \sum_i \max \left( \max(0, f(z)_i - g(x)_i)^2, \max(0, f(z)_i - g(y)_i)^2 \right) \\
&= \sum_i \max \left( 0, f(z) - \min(g(x)_i, g(y)_i) \right)^2 \\
&= E_c(z \rightarrow (x \wedge y))
\end{align*}
\end{proof}

\end{document}